\documentstyle[twocolumn,graphicx]{jpsj}
\def\runtitle{Ferromagnetic and Triplet-Pairing Instabilities Controlled by Trigonal Distortion of CoO$_6$ Octahedra in Na$_x$CoO$_2 \cdot y$H$_2$O}
\def\runauthor{Masahito {\sc Mochizuki}, Youichi {\sc Yanase} and Masao {\sc Ogata}}
\title{Ferromagnetic and Triplet-Pairing Instabilities Controlled by Trigonal Distortion of CoO$_6$ Octahedra in Na$_x$CoO$_2 \cdot y$H$_2$O}
\author
{Masahito {\sc Mochizuki}, Youichi {\sc Yanase} and Masao {\sc Ogata}}
\inst
{Department of Physics, University of Tokyo, Hongo Bunkyo-ku, Tokyo 113-0033, Japan}
\recdate{\today}

\abst{
In the recently discovered Co-oxide superconductor Na$_x$CoO$_2 \cdot y$H$_2$O, the edge-shared CoO$_6$ octahedra are trigonally contracted along the $c$-axis in the CoO$_2$-plane. We study how this CoO$_6$ distortion affects the magnetic properties and superconductivity in this compound by analyzing the multiorbital Hubbard model using the fluctuation-exchange approximation. It is shown that through generating the trigonal crystal field, the distortion pushes the Co ${e'}_g$ bands up and consequently gives rise to the hole-pocket Fermi surfaces, which have been predicted in the band calculations. As the distortion increases, the hole pockets are enlarged and the ferromagnetic fluctuation as well as the pairing instability increases, which is in good agreement with recent NQR results.}

\kword
{Co-oxide superconductor, hole-pocket Fermi surfaces, ferromagnetic fluctuation, triplet pairing}
\begin{document}
\sloppy
\maketitle
The recently discovered Co-oxide superconductor Na$_x$CoO$_2 \cdot y$H$_2$O~\cite{Takada03} has attracted great interest. One of the interesting points is that the unconventional superconductivity due to an electron-correlation mechanism, in particular, the spin-fluctuation-mediated mechanism, is expected in this material. Although there are some controversies, NMR/NQR~\cite{Kato03,Ishida03,Fujimoto03} and $\mu$SR~\cite{Higemoto04,Uemura04} measurements have shown evidence for non-$s$-wave pairing. Various experiments on the nonhydrate compounds show characteristic behaviors of strongly correlated electron systems.~\cite{YWang04,HBYang03} Thus, the magnetic properties have been intensively studied since we recognized that understanding them is essential for elucidating the pairing mechanism. A ferromagnetic (FM) spin fluctuation was predicted in the LSDA calculation~\cite{Singh03} and was claimeded based on NQR experiments by several groups.~\cite{Kato03,Ishida03,Kobayashi03a} In contrast, some groups observed Curie-Weiss behavior of 1/$T_1$~\cite{Fujimoto03} and a temperature-independent Knight shift~\cite{Ning04}, which is evidence against the FM fluctuation. The magnetic properties of this compound are still controversial.

Quite recently, NQR measurements performed by Ihara et al. showed that there is a correlation between the magnetic fluctuation and NQR frequency $\nu_Q$ arising from the $\pm5/2\leftrightarrow\pm7/2$ transition.~\cite{Ihara04a,Ihara04b} They found that the higher-$\nu_Q$ sample has a stronger magnetic fluctuation and a higher superconducting transition temperature $T_c$. In addition, a weak magnetic order was found in a sample with the highest $\nu_Q$. They proposed that the magnetic fluctuation and $T_c$ have a correlation with the distortion of the CoO$_6$ octahedra along the $c$-axis since $\nu_Q$ is expected to scale with the CoO$_6$ distortion from the cubic symmetry. They also proposed that superconductivity occurs in the vicinity of magnetic order, suggesting a spin-fluctuation-mediated pairing. A relationship between $T_c$ and the $c$-axis parameter has also been pointed out by several groups.~\cite{Lynn03,Sakurai04}

Another interesting point is orbital degeneracy. This compound has quasi-two-dimensional CoO$_2$ planes with edge-shared networks of CoO$_6$ octahedra, in which the Co ions form a triangular lattice. Band calculations showed that the Fermi surface is constructed from two bands consisting of three Co $t_{2g}$ orbitals.~\cite{Singh00} Concretely, it was shown that among these three $t_{2g}$ orbitals, the singlet $a_{1g}$ orbital forms a large cylindrical Fermi surface around the $\Gamma$-point and six small hole pockets are constructed from the doublet ${e'}_g$ orbitals near the K-points. Considering these aspects, we have recently proposed that the Na$_x$CoO$_2$ system should be described by a multiorbital model~\cite{Mochizuki04,Yanase04,Indergand05}; most of the previous works are based on single-band models.~\cite{Ogata03,Ikeda04,YTanaka04,Kuroki04a,TWatanabe04} In the studies reported in refs.~\citen{Mochizuki04} and \citen{Yanase04}, we analyzed a multiorbital Hubbard model by using the fluctuation-exchange (FLEX) approximation and perturbation theories, and found several important aspects which had not been expected in the single-band theories. These analyses showed that when the system has the six hole pockets predicted in the band calculations, the interorbital Hund's-rule coupling induces FM fluctuation and this FM fluctuation favors the spin-triplet pairings with $f_{y(y^2-3x^2)}$-wave and $p$-wave symmetries. We discussed that for the FM fluctuation as well as the triplet superconductivity, the presence of ${e'}_g$ hole pockets is crucially important, which was also pointed out by Kuroki et al.~\cite{Kuroki04a} and by Yata et al.~\cite{Yata04}.

\begin{figure}[tdp]
\includegraphics[scale=0.43]{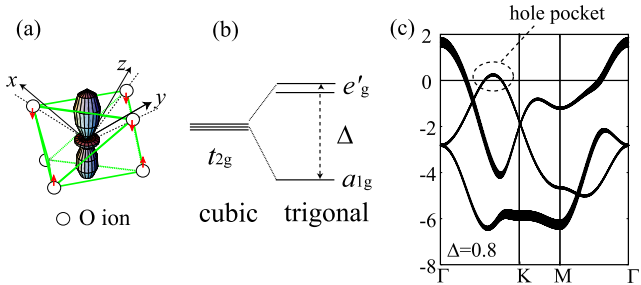}
\caption{(a) Trigonal distortion of CoO$_6$ octahedron and $a_{1g}$ orbital. Arrows indicate shifts of O ions. (b) $a_{1g}$-${e'}_g$ splitting of $t_{2g}$ degeneracy due to trigonal crystal field. (c) Band dispersions calculated from noninteracting Hamiltonian $H_{\rm kin.}+H_{\rm cry.}$. The parameters are chosen to be $t_1=0.38$, $t_2=0.20$, $t_3=1$, $t_4=0.12$, $t_5=-0.15$, $t_6=-0.06$, $t_7=0.12$, $t_8=0.12$, $t_9=-0.30$, and $\Delta=0.8$. The electron number is 5.55 per Co site. The linewidth of each band in (a) is proportional to the amount of the $a_{1g}$-orbital component.}
\label{FIG01}
\end{figure}

In this letter, we study the results of the NQR measurements preformed by Ihara et al.~\cite{Ihara04a,Ihara04b} by extending our previous study on the multiorbital Hubbard model in the FLEX approximation. We show that the results of Ihara et al. can be clearly explained if we consider the hole-pocket Fermi surfaces as well as the Co $t_{2g}$ orbital degrees of freedom. In the CoO$_2$ plane, the CoO$_6$ octahedron is trigonally contracted along the $c$-axis as shown in Fig.~\ref{FIG01}(a). As a result, the O ions generate a trigonal crystal field (TCF) on the Co site, which lifts the local Co $t_{2g}$ degeneracy into the lower singlet $a_{1g}$ and higher doublet ${e'}_g$ orbitals (see Fig.~\ref{FIG01}(b)). First, we show that when the $a_{1g}$-${e'}_g$ splitting becomes larger with increasing CoO$_6$ distortion, the hole pockets appear and become larger near the K-points. As the hole pockets become larger, the FM spin fluctuation becomes stronger and the pairing instability is enhanced, while they are markedly weak without the hole pockets. The successful reproduction of the experimental results seems to support the existence of hole-pocket Fermi surfaces and the pairing mechanism proposed in the previous papers.~\cite{Mochizuki04,Yanase04}

The multiorbital Hubbard Hamiltonian is given by $H=H_{\rm cry.}+H_{\rm kin.}+H_{\rm int.}$ with $H_{\rm cry.}=\sum_{i,m,n,\sigma}D_{mn}d^{\dagger}_{im\sigma}d_{in\sigma}$, $H_{\rm kin.}=\sum_{i,j,m,n,\sigma} t_{ij}^{mn}d^{\dagger}_{im\sigma} d_{jn\sigma}-\mu\sum_{i,m,\sigma} d^{\dagger}_{im\sigma} d_{im\sigma}$, and $H_{\rm int.}=H_{U}+H_{U'}+H_{J_{\rm H}}+H_{J'}$, where $d_{im\sigma}$ ($d^{\dagger}_{im\sigma}$) is the annihilation (creation) operator of an electron with spin $\sigma$(=$\uparrow,\downarrow$) in the orbital $m$ on the Co site $i$. Here, $m$ runs over the $xy$, $yz$, and $zx$ orbitals. 

The first term $H_{\rm cry.}$ expresses the TCF from the O ions. The matrix element $D_{mn}$ is given by $\frac{\Delta}{3}(\delta_{mn}-1)$, which gives the $a_{1g}$-${e'}_g$ splitting of $\Delta$. 
The second term $H_{\rm kin.}$ is a tight-binding Hamiltonian, where the hopping integral $t_{ij}^{mn}$ expresses the electron hopping between the $m$ orbital on the $i$ site and the $n$ orbital on the $j$ site. We choose the lattice constant as a unit length and denote the unit vector as $\vec {a_1}$=($\sqrt{3}$/2,$-$1/2) and $\vec {a_2}$=(0,1). We consider up to the third nearest-neighbor hoppings between orbitals. In the $k$-space, $H_{\rm kin.}$ is rewritten as
$H_{\rm kin.}=\sum_{{\bf k},m,n,\sigma}\epsilon^{mn}_{\bf k}
      d^{\dagger}_{{\bf k}m\sigma} d_{{\bf k}n\sigma}
     -\mu\sum_{{\bf k},m,\sigma}
      d^{\dagger}_{{\bf k}m\sigma} d_{{\bf k}m\sigma}$
with
$\epsilon^{\gamma\gamma}_{\bf k}=2t_1\cos k^{\gamma\gamma}_a + 2t_2\left[\cos k^{\gamma\gamma}_b + \cos (k^{\gamma\gamma}_a+k^{\gamma\gamma}_b)\right] + 2t_4\left[\cos(2k^{\gamma\gamma}_a+k^{\gamma\gamma}_b) + \cos(k^{\gamma\gamma}_a-k^{\gamma\gamma}_b)\right] + 2t_5\cos 2k^{\gamma\gamma}_a$, and $\epsilon^{\gamma\gamma'}_{\bf k}=2t_3\cos k^{\gamma\gamma'}_b + 2t_6\cos 2k^{\gamma\gamma'}_b + 2t_7 \cos(k^{\gamma\gamma'}_a+2k^{\gamma\gamma'}_b) + 2t_8 \cos(k^{\gamma\gamma'}_a-k^{\gamma\gamma'}_b) + 2t_9 \cos(2k^{\gamma\gamma'}_a+k^{\gamma\gamma'}_b)$.
Here, $\gamma$ and $\gamma'$ represent $xy$, $yz$ and $zx$ orbitals and 
$k^{xy,xy}_a=k^{xy,zx}_a=k_1$, $k^{xy,xy}_b=k^{xy,zx}_b=k_2$, 
$k^{yz,yz}_a=k^{xy,yz}_a=k_2$, $k^{yz,yz}_b=k^{xy,yz}_b=-(k_1+k_2)$, 
$k^{zx,zx}_a=k^{yz,zx}_a=-(k_1+k_2)$, and $k^{zx,zx}_b=k^{yz,zx}_b=k_1$, 
respectively. $k_1=\sqrt{3}/2k_x-1/2k_y$ and $k_2=k_y$ are the components of the wave vector $\vec k$ of the triangular lattice spanned by $\vec {a_1}$ and $\vec {a_2}$, respectively. In Fig.~\ref{FIG01}(c), we show the band dispersions obtained from $H_{\rm kin.}+H_{\rm cry.}$ with $t_1=0.38$, $t_2=0.20$, $t_3=1$, $t_4=0.12$, $t_5=-0.15$, $t_6=-0.06$, $t_7=0.12$, $t_8=0.12$, $t_9=-0.30$, and $\Delta=0.8$. In the following, we use these parameters and $t_3=1$ as the energy unit. The LDA results are well reproduced, particularly near the Fermi level.~\cite{Mochizuki04,Singh00} The linewidth of each band is proportional to the amount of the $a_{1g}$-orbital component, which shows that the hole pockets have an ${e'}_g$-orbital character. Here, we note that the dispersions far below the Fermi level are relatively different from those of LDA. This may imply that it is necessary to consider the electron hoppings beyond the third nearest neighbors or the O $2p$ orbital components to reproduce the precise structure of the LDA result. However, we have confirmed that the results in this paper do not depend on details of the band structure far below the Fermi level.

The last term in $H$, $H_{\rm int.}$, represents the on-site $d$-$d$ Coulomb interactions, where $H_{U}$ and $H_{U'}$ are the intraorbital and interorbital Coulomb interactions, respectively, and $H_{J_{\rm H}}$ and $H_{J'}$ are the Hund's-rule coupling and the pair-hopping interactions, respectively. These interactions are expressed using Kanamori parameters, $U$, $U'$, $J_{\rm H}$, and $J'$, which satisfy the relations $U'=U-2J_{\rm H}$ and $J_{\rm H}=J'$. The value of $U$ has been estimated as 3-5.5 eV by photoemission spectroscopy,~\cite{Chainani03} and the value of $J_{\rm H}$ for the Co$^{3+}$ ion is 0.84 eV. Thus, the ratio $J_{\rm H}/U$, which gives the strength of Hund's-rule coupling, is 0.15-0.28 in this compound.

We analyze this multiorbital Hubbard model by extending the FLEX approximation to the multiorbital case.~\cite{Mochizuki04,Takimoto04} In the present three-orbital case, the Green's function $\hat{G}$, the noninteracting Green's function $\hat{G}^{(0)}$ and the self-energy $\hat{\Sigma}$ are expressed in 3$\times$3-matrix form corresponding to the $xy$, $yz$, and $zx$ orbitals.~\cite{Mochizuki04} Calculations are numerically carried out with 64$\times$64 $k$-meshes in the first Brillouin zone and 1024 Matsubara frequencies. The value of $U$ is fixed at 8.0, and temperature $T$ is fixed at 0.02.

\begin{figure}[tdp]
\includegraphics[scale=0.5]{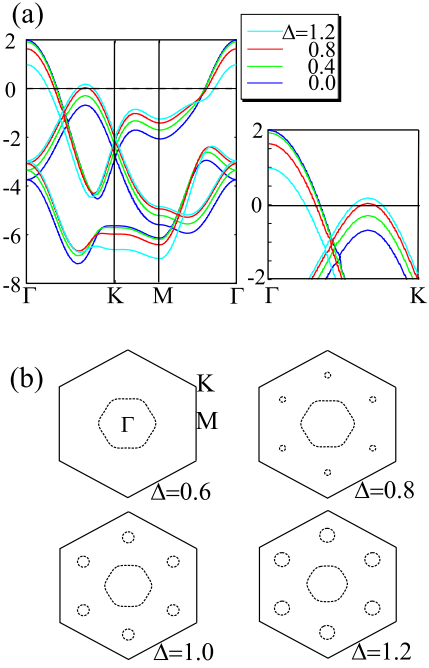}
\caption{(a) Band dispersions and (b) Fermi surfaces for various $\Delta$. The parameters are fixed at $U=8.0$, $J_{\rm H}/U=0.18$ and $T=0.02$.}
\label{FIG02}
\end{figure}
First, we discuss the change of the band structure when the trigonal CoO$_6$ distortion is increased. Since the TCF pushes the ${e'}_g$ level up, we naively expect that the hole-pocket Fermi surfaces would appear as TCF splitting $\Delta$ is increased. In Fig.~\ref{FIG02}, we show the band dispersions (a) and Fermi surfaces (b) in the case of $J_{\rm H}/U=$0.18 and $T=0.02$ for various $\Delta$ values obtained in the FLEX approximation. They are calculated by diagonalizing a matrix whose component is $\epsilon^{mn}_{\bf k}+\Sigma_{mn}({\bf k},\omega=0)$. Here, the self-energy is given by 
\begin{equation}
 \Sigma_{mn}(k)=\frac{T}{N}\sum_{q}\sum_{\mu\nu} V_{\mu m,\nu n}(q)G_{\mu\nu}(k-q)
\end{equation}
with effective interaction $V_{\mu m,\nu n}(q)$ in RPA-type bubble and ladder diagrams.~\cite{Mochizuki04} Figure~\ref{FIG02} actually shows that the ${e'}_g$ bands are gradually pushed up with increasing $\Delta$. As a result, when $\Delta \ge$0.8, the ${e'}_g$ bands intersect the Fermi level and form the six hole pockets near the K-points. With further increasing $\Delta$, the areas of the hole pockets become larger.

\begin{figure}[tdp]
\includegraphics[scale=0.6]{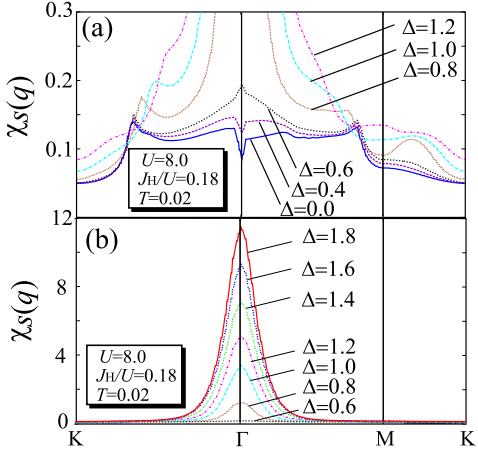}
\caption{(a) Spin susceptibility $\chi^{\rm s}({\bf q})$ for 0$<\Delta<$1.2 and (b) $\chi^{\rm s}({\bf q})$ for 0.6$<\Delta<$1.8. The parameters are fixed at $U=8.0$, $J_{\rm H}/U$=0.18, and $T=$0.02.}
\label{FIG03}
\end{figure}
We next discuss the correlation between the spin fluctuation and the magnitude of the trigonal distortion. Figure~\ref{FIG03} shows the spin susceptibility $\chi^{\rm s}({\bf q})=\chi^{\rm s}({\bf q},\omega=0)$ for various values of $\Delta$ at $J_{\rm H}/U$=0.18 and $T=$0.02. Here, $\chi^{\rm s}(q)$ is given by $\chi^{\rm s}(q)=\sum_{mn} \chi^{\rm s}_{mm,nn}(q)$ with $\hat{\chi}^{\rm s}(q)=[\hat{I}- \hat{\chi}^0(q)\hat{U}^{\rm s}]^{-1} \hat{\chi}^0(q)$ and $\hat{U}^{\rm s}$ is a 6$\times$6 matrix representing the interactions $U$, $U'$, $J_{\rm H}$, and $J'$.~\cite{Mochizuki04} Figure~\ref{FIG03}(a) shows that the nature of spin fluctuation is completely different between the case with hole pockets ($\Delta>0.8$) and the case without hole pockets ($\Delta<0.6$). $\chi^{\rm s}({\bf q})$ for $\Delta>0.8$ has an enhanced peak structure at the $\Gamma$-point indicating the dominant FM fluctuation. This FM fluctuation rapidly increases with increasing $\Delta$ as shown in Fig.~\ref{FIG03}(b). On the other hand, without hole pockets, $\chi^{\rm s}({\bf q})$ does not have any remarkable structure except for very tiny peaks caused by the weak nesting of the $a_{1g}$ Fermi surface. Since the Hund's-rule coupling only works between two electrons on different orbitals, the coupling cannot generate the FM fluctuation when the system only has the $a_{1g}$ Fermi surface. These results suggest that the existence of hole pockets is essential for the appearance of the FM fluctuation. 

\begin{figure}[tdp]
\includegraphics[scale=0.4]{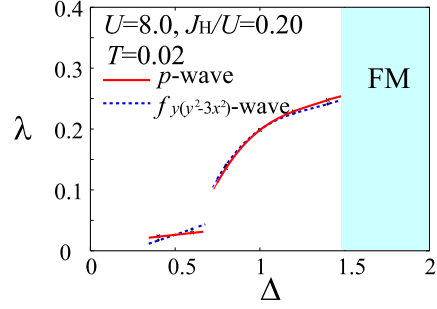}
\caption{Eigenvalues of Eliashberg equation for $f_{y(y^2-3x^2)}$-wave and $p$-wave pairing at $U=8.0$, $J_{\rm H}/U$=0.20, and $T=$0.02 as functions of TCF splitting $\Delta$.}
\label{FIG04}
\end{figure}
To discuss the nature of superconductivity, we solve the following Eliashberg equation:
\begin{eqnarray}
    \lambda \phi_{mn}(k)&=&-\frac{T}{N}\sum_{q}\sum_{\alpha\beta}\sum_{\mu\nu}
    \Gamma^{\eta}_{\alpha m,n\beta}(q) \phi_{\mu\nu}(k-q) \nonumber\\
    &\times& G_{\alpha\mu}(k-q) \hspace{1mm} G_{\beta\nu}(q-k).
\end{eqnarray}
The singlet and triplet pairing interactions $\Gamma^{\rm s}$  and $\Gamma^{\rm t}$ have 9$\times$9-matrix forms as
%
$\hat{\Gamma}^{\rm s}(q)=\frac{3}{2}\hat{U}^{\rm s}\hat{\chi}^{\rm s}(q)\hat{U}^{\rm s}-\frac{1}{2}\hat{U}^{\rm c}\hat{\chi}^{\rm c}(q)\hat{U}^{\rm c}+\frac{1}{2}(\hat{U}^{\rm s}+\hat{U}^{\rm c})$ and $\hat{\Gamma}^{\rm t}(q)=-\frac{1}{2}\hat{U}^{\rm s}\hat{\chi}^{\rm s}(q)\hat{U}^{\rm s}-\frac{1}{2}\hat{U}^{\rm c}\hat{\chi}^{\rm c}(q)\hat{U}^{\rm c}+\frac{1}{2}(\hat{U}^{\rm s}+\hat{U}^{\rm c})$.
Here, $\hat{\chi}^{\rm c}(q)=[\hat{I}+ \hat{\chi}^0(q)\hat{U}^{\rm c}]^{-1} \hat{\chi}^0(q)$ and $\hat{U}^{\rm c}$ is a 6$\times$6 matrix representing the on-site Coulomb interactions.~\cite{Mochizuki04} The eigenvalue $\lambda$ is a measure of the dominant pairing symmetry, and becomes unity at $T=T_{\rm c}$. 

In Figure~\ref{FIG04}, we plot the eigenvalue of the Eliashberg equation ($\lambda$) for both $f_{y(y^2-3x^2)}$-wave and $p$-wave states as functions of the TCF splitting $\Delta$. As discussed previously,~\cite{Mochizuki04,Yanase04} the triplet $f_{y(y^2-3x^2)}$-wave or $p$-wave pairing, which has the large gap amplitude on the ${e'}_g$ hole pockets, has the largest $\lambda$ and the other pairing instabilities are markedly weak. We can see that $\lambda$ for the $f_{y(y^2-3x^2)}$-wave state and that for the $p$-wave state are markedly small in the region $\Delta<0.6$, which indicates that without the hole pockets, the pairing instability is strongly suppressed. From $\Delta\sim0.7$, $\lambda$ rapidly increases as $\Delta$ increases. For $\Delta$ larger than $\sim$1.4, the system exhibits FM ordering. Here, the FM phase transition is identified by ${\rm Min}\{ {\rm det}[\hat{I}- \hat{\chi}^0(q)\hat{U}^{\rm s}] \}=2\times10^{-3}$. This behavior is quite consistent with the NQR result, which suggests the increase of $T_c$ with increasing CoO$_6$ distortion and is consistent with the finding of weak magnetic ordering in a sample with more significant distortion.~\cite{Ihara04a,Ihara04b}

In summary, motivated by the recent NQR experiments,~\cite{Ihara04a,Ihara04b} we studied how the CoO$_6$ distortion along the $c$-axis affects the magnetic fluctuation and the pairing instability in Na$_x$CoO$_2 \cdot y$H$_2$O on the basis of the FLEX analysis of the multiorbital Hubbard model. 
We showed that through generating TCF, the CoO$_6$ distortion gives rise to the hole-pocket Fermi surfaces predicted in the LDA calculation by pushing the ${e'}_g$ bands up. The enhanced FM spin fluctuation is realized when the system has the hole pockets, and the magnetic and pairing instabilities are enhanced as the distortion increases. These results are in good agreement with the experiments and, thus, seem to support the validity of our theoretical proposal about the triplet pairing being mediated by the Hund's-rule-coupling-induced FM fluctuation, which assumed the existence of the hole-pocket Fermi surfaces. Further, the results suggest a possible controllability of $T_c$ and magnetism in this compound via the application of pressure along the $c$-axis. The pressure may possibly induce superconductivity even in the nonhydrate systems.

Note that in angle-resolved photoemission spectroscopy (ARPES) studies on nonhydrate compounds, the hole-pocket Fermi surfaces have not been observed thus far.~\cite{HBYang03,Valla02,Hasan03} This may due to the surface sensitivity of the CoO$_6$ distortion. Our calculations show that the trigonal crystal fields due to the contracted CoO$_6$ octahedra generate the hole pockets by pushing the ${e'}_g$ bands up. The contraction may be relaxed at the surface resulting in the vanishing of hole pockets. Although the existence pf hole pockets has not been confirmed experimentally at present, our results strongly support their presence.

Within our model, the triplet $f_{y(y^2-3x^2)}$-wave and $p$-wave pairings are favorable. To elucidate the pairing state, several groups performed Knight-shift measurements by NMR~\cite{Kato03,Kobayashi03a,Ishida04,Michioka04} and $\mu$SR~\cite{Higemoto04}. Some of these groups reported a constant Knight shift below $T_c$.~\cite{Kato03,Higemoto04,Michioka04} This could be evidence for the triplet pairing. On the other hand, other groups reported a decreasing Knight shift below $T_c$, which suggests the singlet pairing or triplet pairing whose $d$-vector is fixed in the $ab$-plane.~\cite{Kobayashi03a,Ishida04} These two results seem to be contradictory at first sight. However, they may be understood from difference in the experimental conditions. In fact, the former measurements were performed under relatively high magnetic fields, while the latter ones were performed under rather low fields. These two results may suggest the possible existence of multiple superconducting phases under the magnetic field, and the $d$-vector may be directed along the $c$-axis in the high-field phase, while in the low-field phase it may be directed in the $ab$-plane. Quite recently, it was reported that the Knight shift decreases even when the magnetic field is parallel to the $c$-axis.~\cite{Kobayashi04}. However, the decrease is much smaller than that expected from the $c$-axis magnetic susceptibility. Complete understanding of the NMR results requires further theoretical studies on the direction of the $d$-vector by considering the effects of the spin-orbit interaction and magnetic fields.

We thank K. Ishida, M. Takigawa, Y. Ono, K. Yoshimura, G.-q. Zheng, Y. Kobayashi, M. Sato, R. Kadono, K. Kuroki, and G. Baskaran for valuable discussions. This work is supported by Grant-in-Aid for Scientific Research from MEXT.

\end{document}